\documentclass[a4paper,11pt]{article}
\usepackage{pos}
\usepackage{amsmath}
\usepackage{amscd}
\usepackage{amsfonts}
\usepackage{mathrsfs}
\usepackage{enumerate}
\usepackage[boxsize=0.8ex]{ytableau} 
\usepackage{youngtab}
\usepackage{graphicx}
\usepackage{url} 
\usepackage{color}
\usepackage{braket}

\usepackage{comment}

\def\={\stackrel{\bullet}{=}}

\def\({\left(}
\def\){\right)}
\def\[{\left[}
\def\]{\right]}

\def\i{\mathrm{i}}

\def\KS{\text{KS}}

\def\W{\text{W}}

\title{The Axial Charge in Hilbert Space and the Role in Chiral Gauge Theories}

\author*{Tatsuya~Yamaoka}

\affiliation[]{Department of Physics, the University of Osaka, Toyonaka, Osaka 560-0043, Japan}

\emailAdd{t\_yamaoka@het.phys.sci.osaka-u.ac.jp}

\abstract{
We investigate the Hamiltonian formulation of 1+1-dimensional
staggered fermions and reconstruct the vector and axial charge operators,
originally identified by Arkya Chatterjee \textit{et al.},
within the Wilson fermion formalism.
These operators commute with the Hamiltonian and reduce,
in the continuum limit, to the generators of the vector and axial
$\mathrm{U}(1)$ symmetries.
A notable feature of the axial charge operator is that it acts locally on operators
and possesses quantized eigenvalues.
Its eigenstates can therefore be interpreted as fermion states
with well-defined integer chirality, analogous to those in the continuum theory.
This structure enables the formulation of a gauge theory
in which the axial $\mathrm{U}(1)_A$ symmetry is promoted to a gauge symmetry.
We construct a Hamiltonian in terms of the eigenstates of the axial charge operator,
thereby preserving exact axial symmetry on the lattice
while recovering vector symmetry in the continuum limit.
As applications, we study the implementation of the Symmetric Mass Generation (SMG) mechanism
in the 3-4-5-0 models.
Our framework admits symmetry-preserving interaction terms
with quantized chiral charges,
although further numerical investigation is required to confirm
the realization of the SMG mechanism in interacting systems.

OU-HET-1305
}

\FullConference{The 42nd International Symposium on Lattice Field Theory (LATTICE2025)\\
2-8 November 2025\\
Tata Institute of Fundamental Research, Mumbai, India\\}

\begin{document}
\maketitle

\section{Introduction}
\label{sec:Introduction}
The lattice formulation of chiral gauge theories remains a longstanding challenge in quantum field theory. 
The primary obstruction is the Nielsen--Ninomiya no-go theorem~\cite{Nielsen:1980rz,Nielsen:1981hk}, 
which states that any local, Hermitian, and translationally invariant lattice fermion action 
necessarily yields fermion doublers as a consequence of the periodicity of the Brillouin zone. 
Removing these doublers inevitably breaks chiral symmetry, 
making the nonperturbative formulation of chiral gauge theories highly nontrivial.

In the path-integral formalism, substantial progress has been achieved through the overlap fermion construction~\cite{Neuberger:1998wv}, 
which is based on the Ginsparg--Wilson relation~\cite{Ginsparg:1981bj}. 
Overlap fermions preserve a modified chiral symmetry at finite lattice spacing 
and allow for a controlled treatment of chiral properties on the lattice. 
However, the corresponding Hamiltonian formulation of lattice chiral symmetry 
is still under active investigation~\cite{Chatterjee:2024gje,Onogi:2025xir,Hidaka:2025ram,Singh:2025sye,Aoki:2025vtp,Misumi:2025yjf,Gioia:2025bhl,Thorngren:2026ydw,Seifnashri:2026ema}.

Recently, Ref.~\cite{Chatterjee:2024gje} proposed a new formulation of 1+1-dimensional staggered fermions 
in which both vector and axial charge operators are local, commute with the Hamiltonian, 
and possess quantized eigenvalues. 
These operators generate independent $\mathrm{U}(1)$ symmetries 
that can in principle be gauged separately. 
The noncommutativity between the vector and axial charges 
encodes the chiral anomaly within the Hamiltonian framework.

In this work, we revisit this construction from the viewpoint of Wilson fermions. 
The result was already published in Ref~\cite{Yamaoka:2025sdm}. 
We demonstrate that, in 1+1 dimensions, 
the staggered fermion Hamiltonian can be smoothly deformed into the Wilson fermion Hamiltonian. 
Using this equivalence, we reinterpret the quantized axial charge in terms of Wilson fermion variables. 
The eigenstates of the axial charge are constructed as linear combinations of 
positive-energy creation and negative-energy annihilation operators, 
leading to a Hamiltonian that contains particle-number nonconserving terms. 
Nevertheless, in the continuum limit, the theory still admits conserved 
vector and axial $\mathrm{U}(1)$ charges that remain noncommutative, 
thereby avoiding contradiction with the Nielsen--Ninomiya theorem.

The resulting Hamiltonian formulation is particularly useful for constructing 
chiral gauge theories via the symmetric mass generation (SMG) mechanism~\cite{Wang:2013yta,Wang:2018ugf}. 
SMG provides a way to gap fermions without fermion bilinear mass terms 
while preserving symmetries, provided that the underlying theory is anomaly-free. 
Motivated by this perspective, we reconsider the 3-4-5-0 chiral model 
and clarify the role of the conserved charges in determining whether anomaly cancellation 
occurs nontrivially or only in a trivial lattice sense. 
Based on this analysis, we reconstruct the model within the present framework.

This paper is organized as follows. 
In Sec.~\ref{sec:Equivalence-Stg-Wilson}, we first demonstrate the equivalence between staggered and Wilson fermion Hamiltonians in 1+1 dimensions. 
We then review the quantized lattice charge operators and reformulate the axial charge 
in terms of Wilson fermions in Sec.~\ref{sec:Axial-charge-Staggered-fermions}. 
In Sec.~\ref{sec:Eigenstates-Axial-charge}, 
we redefine the axial charge operator $Q_A$ using Wilson fermions and construct the Hamiltonian in terms of the fields with definite charge of $Q_A$. 
We verify that this Hamiltonian still supports two conserved charges corresponding to vector and axial $\mathrm{U}(1)$ symmetries in the continuum limit.
Next, in Sec.~\ref{sec:SMG} we analyze the conserved symmetries in the continuum limit 
and discuss the application to SMG constructions, 
the 3-4-5-0 models. Sec.~\ref{sec:Conclusion} is devoted to our conclusion.

\section{Equivalence of Staggered and Wilson Fermions in $1+1$ Dimensions}
\label{sec:Equivalence-Stg-Wilson}

In this section, we demonstrate that, in 1+1 dimensions, 
the Hamiltonian of massless free staggered fermions 
is equivalent to that of Wilson fermions~\cite{Hayata:2023skf}. 
This equivalence provides the basis for reinterpreting 
the quantized axial charge in the Wilson fermion framework.

We begin with the Hamiltonian of the 1+1-dimensional staggered fermion
\cite{Kogut:1974ag,Banks:1975gq,Susskind:1976jm,Catterall:2025vrx},
\begin{align}
    \label{eq:Def:Staggered-Hamiltonian}
    H_{\text{KS}} 
    &= \i \sum_{j=1}^{2N} \(c_j^\dagger c_{j+1} + c_j c_{j+1}^\dagger\) \notag\\
    &= \frac{\i}{2} \sum_{j=1}^{2N} \(a_{j} a_{j+1} + b_{j}b_{j+1}\) \, ,
\end{align}
where $c_j$ is a single-component complex fermion satisfying
\begin{align}
    \{ c_j , c_{j'}^{\dagger} \} = \delta_{j,j'} \, .
\end{align}
Decomposing $c_j$ into two Majorana fermions,
\begin{align}
    c_j = \frac{1}{2} \( a_j + b_j \) \, ,
\end{align}
with
\begin{align}
    a_j = a_j^\dagger, \quad b_j = b_j^\dagger, \quad \{a_{j}, a_{j'}\} = \{b_{j}, b_{j'}\} = 2 \delta_{jj'} \, ,
\end{align}
one sees that Eq.~\eqref{eq:Def:Staggered-Hamiltonian} 
describes a single massless Dirac fermion in the continuum limit.

Introducing a two-component field
\begin{align}
    \psi_j = \begin{pmatrix}
        \psi_{L,j} \\
        \psi_{R,j}
    \end{pmatrix}=\begin{pmatrix}
        c_{2j} \\
        \i c_{2j+1}
    \end{pmatrix} \, ,
\end{align}
and defining
\begin{align}
    \gamma^0 = \sigma^1 \, , \quad \gamma^1 = -\sigma^3 \, , \quad \gamma^5 = \sigma^2 \, ,
\end{align}
the Hamiltonian can be rewritten as
\begin{align}
    H_{\text{KS}} &= \sum_{j=1}^{N} \psi^{\dagger}_j \gamma^0 \left[ \gamma^1 \frac{1}{2}(\nabla + \nabla^{\dagger}) - \frac{1}{2} \nabla^{\dagger} \nabla \right] \psi_j \notag \\
    &= \sum_{j=1}^{N} \psi^{\dagger}_j \gamma^0 \mathcal{D}_{W}(0) \psi_j \equiv H_{\text{W}} \ ,
\end{align}
where the lattice derivatives are defined by
\begin{align}
    \nabla \psi_j &= \psi_{j+1} - \psi_j \, , \quad \nabla^\dagger \psi_j = \psi_j - \psi_{j-1} \, , \notag \\
    \nabla^\dagger \nabla \psi_j &= \psi_{j+1} + \psi_{j-1} - 2\psi_j \, ,
\end{align}
and $\mathcal{D}_W(m)$ denotes the Wilson--Dirac operator
\begin{align}
    \mathcal{D}_{W}(m) = \frac{1}{2}\{\gamma^1(\nabla + \nabla^\dagger) - \nabla^\dagger \nabla \} - m \, .
\end{align}
Thus, in 1+1 dimensions, the staggered fermion Hamiltonian 
is smoothly connected to the Wilson fermion Hamiltonian 
in the massless free limit.

\section{Conserved Charges in the Hamiltonian of Staggered Fermions}
\label{sec:Axial-charge-Staggered-fermions}

A remarkable property of the 1+1-dimensional staggered fermion Hamiltonian 
with a finite lattice Hilbert space 
is the existence of two conserved charge operators~\cite{Chatterjee:2024gje}, 
denoted by $Q_V$ and $Q_A$. 
In the continuum limit, these correspond to the vector and axial $\mathrm{U}(1)$ symmetries, respectively.

Both charges are local operators and commute with the Hamiltonian,
\begin{align}
    [H_\KS, Q_V] = [H_\KS, Q_A] = 0 \ .
\end{align}
They are explicitly given by
\begin{align}
    Q_V 
    &= \sum_{j=1}^{2N} 
    \(c_j^\dagger c_j - \frac{1}{2}\)
    = \frac{\i}{2} \sum_{j=1}^{2N} a_j b_j 
    \equiv \sum_{j=1}^{2N} q_j^V \ , \\
    Q_A 
    &= \frac{1}{2} \sum_{j=1}^{2N} 
    \(c_j + c_j^\dagger\)
    \(c_{j+1} - c_{j+1}^\dagger\)
    = \frac{\i}{2} \sum_{j=1}^{2N} a_j b_{j+1} 
    \equiv \sum_{j=1}^{2N} q_{j+\frac{1}{2}}^A \ .
\end{align}
Importantly, both $Q_V$ and $Q_A$ have quantized eigenvalues 
and can be gauged independently at finite lattice spacing.

However, on a finite lattice they do not commute with each other,
\begin{align}
    [Q_V, Q_A]
    &= - \sum_{j=1}^{2N}
    \(c_j c_{j+1} + c_j^\dagger c_{j+1}^\dagger\)= \i G_1 \neq 0 \ ,
\end{align}
where $G_1$ is one of the generators of the Onsager algebra . 
This non-Abelian algebraic structure disappears in the continuum limit, 
where the commutator vanishes. 

The noncommutativity between $Q_V$ and $Q_A$ 
encodes the mixed anomaly between vector and axial $\mathrm{U}(1)$ symmetries 
within the Hamiltonian formalism, 
in a manner consistent with the Nielsen--Ninomiya theorem.

\section{Eigenstates of the Axial Charge and the Hamiltonian}
\label{sec:Eigenstates-Axial-charge}

Using the equivalence between staggered and Wilson fermions 
established in the previous section, 
we now reformulate the conserved charges $Q_V$ and $Q_A$ 
in the Wilson fermion framework. 
In particular, the axial charge operator $Q_A$ 
is an on-site symmetry with quantized eigenvalues in coordinate space. 
Its eigenstates therefore admit an interpretation 
as fermions with integer chirality, 
analogous to Weyl fermions in the continuum theory. 
This structure enables the construction of lattice gauge theories 
with a gauged $\mathrm{U}(1)_A$ symmetry.

To make this explicit, we construct fermionic operators 
that diagonalize $Q_A$. 
The following operators are eigenstates of $Q_A$ 
with eigenvalues $\pm 1$:
\begin{align}
    \label{eq:Def:Axial-fermion-Coordinate-Space-Pi/4-dagger}
    \begin{pmatrix}
        \Psi_{L,j}^{\dagger} \\
        \Psi_{R,j}^{\dagger} 
    \end{pmatrix}
    &= \frac{1}{2\sqrt{2}} 
    \begin{pmatrix}
        -2c_{2j}^\dagger 
        + (c_{2j+1} - c_{2j+1}^\dagger) 
        - (c_{2j-1} + c_{2j-1}^\dagger) \\
         2c_{2j}^\dagger 
        + (c_{2j+1} - c_{2j+1}^\dagger) 
        - (c_{2j-1} + c_{2j-1}^\dagger)
    \end{pmatrix}.
\end{align}
They satisfy
\begin{align}
    [Q_A, \Psi_{L,j}^{\dagger}] = \Psi_{L,j}^{\dagger}, 
    \qquad 
    [Q_A, \Psi_{R,j}^{\dagger}] = -\Psi_{R,j}^{\dagger},
\end{align}
and obey canonical anticommutation relations.

In terms of these fields, the Wilson Hamiltonian becomes
\begin{align}
    \label{eq:Hamiltonian-PsiH-position}
    H_{\W} 
    = \i \sum_{j=1}^{N} \biggl[ 
        &\Psi_{L,j}^{\dagger} 
        \frac{1}{2}(\nabla + \nabla^\dagger) 
        \Psi_{L,j}
        - \Psi_{R,j}^{\dagger} 
        \frac{1}{2}(\nabla + \nabla^\dagger) 
        \Psi_{R,j} \notag \\
        & - \Psi_{R,j} 
        \frac{1}{2} \nabla\nabla^\dagger 
        \Psi_{L,j} 
        - \Psi_{L,j}^{\dagger} 
        \frac{1}{2} \nabla\nabla^\dagger 
        \Psi_{R,j}^{\dagger}
    \biggr].
\end{align}
The last line contains particle-number nonconserving terms 
originating from the Wilson term. 
Thus, fermion number is not manifestly conserved in this basis. 
Nevertheless, the Hamiltonian still admits conserved charges 
$Q_V$ and $Q_A$, satisfying
\begin{align}
    [H_\W, Q_V] = [H_\W, Q_A] = 0.
\end{align}

In this basis,
\begin{align}
    Q_A 
    = \sum_{j=1}^{N} 
    (\Psi_{L,j}^{\dagger} \Psi_{L,j} 
    - \Psi_{R,j}^{\dagger} \Psi_{R,j}),
\end{align}
which makes the integer-valued chirality explicit. 
The vector charge $Q_V$ takes a more involved form, 
reflecting the mixing induced by the Wilson term.

A clearer picture emerges in momentum space. 
Expanding the staggered fermion as
\begin{align}
    c_j = \frac{1}{\sqrt{2N}} 
    \sum_k \gamma_k 
    e^{\frac{2\pi i}{2N}kj},
\end{align}
one finds that the Hamiltonian becomes
\begin{align}
    H_\W 
    &= \frac{1}{2} 
    \sum_{- \frac{N}{2} \leq k < \frac{N}{2}} 
    \biggl[
        (\tilde{\Psi}_{L,k}^{\dagger}\tilde{\Psi}_{L,k}
        - \tilde{\Psi}_{R,k}^{\dagger}\tilde{\Psi}_{R,k})
        \sin \frac{2\pi k}{N}
        \notag \\
        &\qquad
        - \i 
        (\tilde{\Psi}_{L,k}^{\dagger}\tilde{\Psi}_{R,-k}^{\dagger}
        - \tilde{\Psi}_{R,-k}\tilde{\Psi}_{L,k})
        \Bigl(1-\cos \frac{2\pi k}{N}\Bigr)
    \biggr].
\end{align}
The second term corresponds to the Wilson term 
and vanishes linearly in momentum near $k=0$.

Taking the continuum limit $N\to\infty$ 
at fixed physical momentum, 
the commutators reduce to
\begin{align}
     [Q_V, \tilde{\Psi}_{\alpha,k}^{\dagger}]
     &=  \tilde{\Psi}_{\alpha,k}^{\dagger}, \\
     [Q_V, \tilde{\Psi}_{\alpha,k}]
     &= -\tilde{\Psi}_{\alpha,k},
\end{align}
showing that $Q_V$ flows to the generator 
of the vector $\mathrm{U}(1)_V$ symmetry, 
while $Q_A$ generates $\mathrm{U}(1)_A$.

Although the Hamiltonian appears to violate 
particle number conservation at finite lattice spacing, 
it preserves two noncommuting conserved charges. 
Because $Q_V$ and $Q_A$ do not commute on a finite lattice 
and their eigenvalues are not simultaneously quantized, 
they cannot be gauged simultaneously. 
This structure ensures consistency with 
the Nielsen--Ninomiya theorem, 
while allowing well-defined Weyl fermions 
with integer chirality over the entire Brillouin zone.

\section{Application to Symmetric Mass Generation}
\label{sec:SMG}

A promising strategy for constructing chiral gauge theories on the lattice 
is the mechanism of symmetric mass generation (SMG)~\cite{Wang:2022ucy,Razamat:2020kyf,Tong:2021phe}. 
SMG allows fermions to acquire a mass gap through multi-fermion interactions 
without introducing any fermion bilinear mass terms, 
while preserving the imposed symmetries. 
Although such interactions are typically irrelevant in weak coupling, 
numerical studies indicate that they can become relevant 
in strong-coupling regimes~\cite{Zeng:2022grc}.

For a fermionic system with symmetry group $G$, 
the realization of SMG requires:
\begin{enumerate}[(i)]
  \item The symmetry $G$ must be anomaly-free.
  \item $G$ must forbid all fermion bilinear mass terms.
  \item The instanton saturation condition must be satisfied.
\end{enumerate}
The third condition becomes essential in the presence of background or dynamical 
$G$ gauge fields with nontrivial topological charge. 
Such backgrounds generate chiral zero modes, 
which must be saturated by interaction terms in order to maintain consistency. 

In Sec.~\ref{sec:Eigenstates-Axial-charge}, 
we constructed Weyl fermions with integer-valued axial charges. 
This structure enables us to gauge the $\mathrm{U}(1)_A$ symmetry 
generated by $Q_A$ and to investigate SMG within this framework.

In the present proceedings, 
we focus on the 3-4-5-0 model~\cite{Wang:2013yta,Wang:2018ugf,Wang:2022fzc,Zeng:2022grc,Onogi:2025tev}. 
A detailed analysis of the $1^4(-1)^4$ model 
has been presented separately in Ref.~\cite{Yamaoka:2025sdm}, 
and will not be repeated here.

\subsection{3-4-5-0 model}
\label{subsec:3450-model}

We begin with four flavors of massless Dirac fermions,
\begin{align}
    \label{eq:Def:3-4-5-0-Hamiltonian}
    H =  \i \sum_{f=1}^{4} \sum_{j=1}^{N} \biggl[
        \Psi_{f,L,j}^{H\dagger} \frac{1}{2}(\nabla + \nabla^\dagger) \Psi_{f,L,j}^{H}
        - \Psi_{f,R,j}^{H\dagger} \frac{1}{2}(\nabla + \nabla^\dagger) \Psi_{f,R,j}^{H}
        - \Psi_{f,R,j}^{H} \frac{1}{2} \nabla \nabla^\dagger \Psi_{f,L,j}^{H}
        - \Psi_{f,L,j}^{H\dagger} \frac{1}{2} \nabla \nabla^\dagger \Psi_{f,R,j}^{H\dagger}
    \biggr].
\end{align}
The goal is to gap out one chirality of each flavor, 
leaving the desired chiral spectrum in the infrared.

The Hamiltonian preserves four $\mathrm{U}(1)$ symmetries 
generated by flavor-resolved vector and axial charges, 
$Q_{V_f}$ and $Q_{A_f}$. 
From these, we define\footnote{
        In principle, the charge assignments of the $\mathrm{U}(1)$ gauge symmetry and the extra $\mathrm{U}(1)$ symmetry in this model can be uniquely determined, 
        both in the lattice and continuum theories, by the condition that anomaly matching is equivalent to the boundary fully gapping condition~\cite{Wang:2013yta,Wang:2018ugf}. 
In our lattice construction, however, the anomaly matching conditions for the $\mathrm{U}(1)$ gauge symmetry and the extra $\mathrm{U}(1)$ symmetry become trivial by construction, i.e., $[Q_{a1}, Q_{a2}] = 0$. As a result, the charge assignments are no longer uniquely fixed by anomaly considerations alone. 
In this work, we therefore choose the charge assignments so that they consistently reproduce those of the 3-4-5-0 model in the continuum limit. 
A discussion of anomaly cancellation in this model can also be found in Ref.~\cite{Li:2024dpq}.
}
\begin{align}
    Q_{a1} &= 3 Q_{A_1} + 4 Q_{A_2} - 5 Q_{A_3}, \\
    Q_{a2} &= 5 Q_{A_2} - 4 Q_{A_3} - 3 Q_{A_4}, \\
    Q_{v1} &= 3 Q_{V_1} + 4 Q_{V_2} - 5 Q_{V_3}, \\
    Q_{v2} &= 5 Q_{V_2} - 4 Q_{V_3} - 3 Q_{V_4}.
\end{align}
These reproduce the charge assignments 
of the continuum 3-4-5-0 model.

At this stage the theory is vector-like, 
but fermion bilinear mass terms are forbidden 
by $\mathrm{U}(1)_{a1}\times \mathrm{U}(1)_{a2}$. 
In our construction, 
the anomaly cancellation within $\mathrm{U}(1)_{a1}\times \mathrm{U}(1)_{a2}$ 
is realized trivially at the lattice level, 
while mixed anomalies between axial and vector symmetries remain:
\begin{align}
    [Q_{a,s}, Q_{v,s'}] \neq 0.
\end{align}
Therefore, to gap out selected chiral modes while preserving 
$\mathrm{U}(1)_{a1}\times \mathrm{U}(1)_{a2}$, 
one must explicitly break the anomalous 
$\mathrm{U}(1)_{v1}\times \mathrm{U}(1)_{v2}$ 
via multi-fermion interactions.

Such interactions can be chosen as
\begin{align}
    \Delta_{H1} &= \sum_j (\Delta_1 + \mathrm{h.c.}), \\
    \Delta_{H2} &= \sum_j (\Delta_2 + \mathrm{h.c.}),
\end{align}
with
\begin{align}
    \Delta_1 &\propto 
    \Psi_{1,L,j}
    (\Psi_{2,L,j}^\dagger \Psi_{2,L,j+1}^\dagger)
    \Psi_{3,R,j}
    (\Psi_{4,R,j} \Psi_{4,R,j+1}), \\
    \Delta_2 &\propto 
    (\Psi_{1,L,j}\Psi_{1,L,j+1})
    \Psi_{2,L,j}
    (\Psi_{3,R,j}^\dagger \Psi_{3,R,j+1}^\dagger)
    \Psi_{4,R,j}.
\end{align}
These interaction terms commute with $Q_{a1}$ and $Q_{a2}$ 
throughout the entire Brillouin zone, 
while explicitly breaking the vector symmetries. 
Moreover, appropriate combinations of $\Delta_1$ and $\Delta_2$ 
generate the required 't~Hooft vertices, 
thereby satisfying the instanton saturation condition.

Whether the intended chiral spectrum is dynamically realized 
requires further numerical investigation. 
Nevertheless, the present construction demonstrates 
that the SMG framework can be consistently implemented 
with a gauged $\mathrm{U}(1)_A$ symmetry 
generated by the quantized axial charge.


\section{Conclusion}
\label{sec:Conclusion}

In this work, we have reconstructed two lattice charge operators,
$Q_V$ and $Q_A$, that commute with the Hamiltonian of the 1+1-dimensional staggered fermion,
by employing the Wilson fermion formalism.
We have shown that these operators reduce, in the continuum limit,
to the generators of the vector $\mathrm{U}(1)_V$ and axial $\mathrm{U}(1)_A$ symmetries, respectively.
Furthermore, we constructed the Hamiltonian explicitly in terms of the eigenstates of $Q_A$,
thereby ensuring the exact preservation of the axial $\mathrm{U}(1)_A$ symmetry on the lattice.

Although the resulting Hamiltonian may appear to break the naive vector
$\mathrm{U}(1)_V$ symmetry at the lattice level,
it is, by construction, compatible with $Q_V$ and therefore reproduces
the correct $\mathrm{U}(1)_V$ symmetry in the continuum limit.
While this observation may appear straightforward,
it is nevertheless nontrivial that the Hamiltonian remains consistent with $Q_V$
even when it is explicitly designed to preserve axial symmetry on the lattice.
A key advantage of maintaining exact $Q_A$ symmetry is that the operator acts locally,
which in turn allows it to be gauged.
This opens the possibility of formulating chiral $\mathrm{U}(1)_A$ gauge theories on the lattice,
at least within certain controlled settings.

As an application, we discussed how the Symmetric Mass Generation (SMG) mechanism
can be implemented in free chiral $\mathrm{U}(1)_A$ gauge theories
(without dynamical gauge fields),
using the 3-4-5-0 model as illustrative examples.
Because the Hamiltonian is formulated in terms of fermions
with well-defined, momentum-independent integer-valued chirality,
the interaction terms required for gapping can be introduced consistently
without violating the symmetries that must be preserved.
In the case of the 3-4-5-0 model, however,
further numerical investigation is necessary
to determine whether the SMG mechanism is genuinely realized in this framework.
Since the Hamiltonian contains Wilson-type terms that generically mix chiralities,
it is highly nontrivial whether one chirality can be fully gapped
while the other remains massless solely through interaction effects.
We leave this important question for future study.

Finally, one of the motivations for adopting the Hamiltonian formulation
is its natural compatibility with quantum simulation platforms.
In particular, our framework may provide a theoretical foundation
for future implementations using ultracold atoms,
which could offer a promising avenue for overcoming the sign problem
in strongly correlated systems~\cite{Zohar:2015hwa}.


\bibliographystyle{JHEP}
\bibliography{reference}

@article{Nielsen:1980rz,
    author = "Nielsen, Holger Bech and Ninomiya, M.",
    editor = "Julve, J. and Ram\'on-Medrano, M.",
    title = "{Absence of Neutrinos on a Lattice. 1. Proof by Homotopy Theory}",
    reportNumber = "RL-80-090",
    doi = "10.1016/0550-3213(82)90011-6",
    journal = "Nucl. Phys. B",
    volume = "185",
    pages = "20",
    year = "1981",
    note = "[Erratum: Nucl.Phys.B 195, 541 (1982)]"
}

@article{Nielsen:1981hk,
    author = "Nielsen, Holger Bech and Ninomiya, M.",
    title = "{No Go Theorem for Regularizing Chiral Fermions}",
    reportNumber = "RL-81-052",
    doi = "10.1016/0370-2693(81)91026-1",
    journal = "Phys. Lett. B",
    volume = "105",
    pages = "219--223",
    year = "1981"
}

@article{Chatterjee:2024gje,
    author = "Chatterjee, Arkya and Pace, Salvatore D. and Shao, Shu-Heng",
    title = "{Quantized Axial Charge of Staggered Fermions and the Chiral Anomaly}",
    eprint = "2409.12220",
    archivePrefix = "arXiv",
    primaryClass = "hep-th",
    reportNumber = "MIT-CTP/5760, YITP-SB-2024-19",
    doi = "10.1103/PhysRevLett.134.021601",
    journal = "Phys. Rev. Lett.",
    volume = "134",
    number = "2",
    pages = "021601",
    year = "2025"
}

@article{Catterall:2025vrx,
    author = "Catterall, Simon and Pradhan, Arnab and Samlodia, Abhishek",
    title = "{Symmetries and Anomalies of Hamiltonian Staggered Fermions}",
    eprint = "2501.10862",
    archivePrefix = "arXiv",
    primaryClass = "hep-lat",
    month = "1",
    year = "2025"
}

@article{Gioia:2025bhl,
    author = "Gioia, Lei and Thorngren, Ryan",
    title = "{Exact Chiral Symmetries of 3+1D Hamiltonian Lattice Fermions}",
    eprint = "2503.07708",
    archivePrefix = "arXiv",
    primaryClass = "cond-mat.str-el",
    month = "3",
    year = "2025"
}

@article{Yamaoka:2025sdm,
    author = "Yamaoka, Tatsuya",
    title = "{Quantized Axial Charge in the Hamiltonian Approach to Wilson Fermions}",
    eprint = "2504.10263",
    archivePrefix = "arXiv",
    primaryClass = "hep-lat",
    reportNumber = "OU-HET 1271",
    month = "4",
    year = "2025"
}

@article{Onogi:2025xir,
    author = "Onogi, Tetsuya and Yamaoka, Tatsuya",
    title = "{Non-singlet conserved charges and anomalies in 3+1 D staggered fermions}",
    eprint = "2509.04906",
    archivePrefix = "arXiv",
    primaryClass = "hep-lat",
    reportNumber = "OU-HET-1287",
    month = "9",
    year = "2025"
}

@article{Aoki:2025vtp,
    author = "Aoki, Shoto and Kikukawa, Yoshio and Takemoto, Toshinari",
    title = "{Chiral Anomaly of Kogut-Susskind Fermion in (3+1)-dimensional Hamiltonian formalism}",
    eprint = "2511.06198",
    archivePrefix = "arXiv",
    primaryClass = "hep-lat",
    reportNumber = "UT-Komaba/25-10, RIKEN-iTHEMS-Report-25",
    month = "11",
    year = "2025"
}

@article{Misumi:2025yjf,
    author = "Misumi, Tatsuhiro",
    title = "{Minimal-doubling and single-Weyl Hamiltonians}",
    eprint = "2512.22609",
    archivePrefix = "arXiv",
    primaryClass = "hep-lat",
    month = "12",
    year = "2025"
}

@article{Hayata:2023skf,
    author = "Hayata, Tomoya and Nakayama, Katsumasa and Yamamoto, Arata",
    title = "{Dynamical chirality production in one dimension}",
    eprint = "2309.08820",
    archivePrefix = "arXiv",
    primaryClass = "hep-lat",
    reportNumber = "RIKEN-iTHEMS-Report-23",
    doi = "10.1103/PhysRevD.109.034501",
    journal = "Phys. Rev. D",
    volume = "109",
    number = "3",
    pages = "034501",
    year = "2024"
}

@article{Hidaka:2025ram,
    author = "Hidaka, Yoshimasa and Yamamoto, Arata",
    title = "{Anomaly of conserved and nonconserved axial charges in Hamiltonian lattice gauge theory}",
    eprint = "2506.01336",
    archivePrefix = "arXiv",
    primaryClass = "hep-lat",
    reportNumber = "RIKEN-iTHEMS-Report-25, YITP-25-84",
    doi = "10.1093/ptep/ptaf114",
    journal = "PTEP",
    volume = "2025",
    number = "9",
    year = "2025"
}

@article{Singh:2025sye,
    author = "Singh, Hersh",
    title = "{Ginsparg-Wilson Hamiltonians with Improved Chiral Symmetry}",
    eprint = "2505.20419",
    archivePrefix = "arXiv",
    primaryClass = "hep-lat",
    reportNumber = "FERMILAB-PUB-25-0330-T",
    month = "5",
    year = "2025"
}

@article{Thorngren:2026ydw,
    author = "Thorngren, Ryan and Preskill, John and Fidkowski, Lukasz",
    title = "{Chiral Lattice Gauge Theories from Symmetry Disentanglers}",
    eprint = "2601.04304",
    archivePrefix = "arXiv",
    primaryClass = "hep-th",
    month = "1",
    year = "2026"
}

@article{Seifnashri:2026ema,
    author = "Seifnashri, Sahand",
    title = "{Exactly Solvable 1+1d Chiral Lattice Gauge Theories}",
    eprint = "2601.14359",
    archivePrefix = "arXiv",
    primaryClass = "hep-th",
    month = "1",
    year = "2026"
}

@article{Kogut:1974ag,
    author = "Kogut, John B. and Susskind, Leonard",
    title = "{Hamiltonian Formulation of Wilson's Lattice Gauge Theories}",
    reportNumber = "Print-74-1186 (CORNELL)",
    doi = "10.1103/PhysRevD.11.395",
    journal = "Phys. Rev. D",
    volume = "11",
    pages = "395--408",
    year = "1975"
}

@article{Banks:1975gq,
    author = "Banks, Tom and Susskind, Leonard and Kogut, John B.",
    title = "{Strong Coupling Calculations of Lattice Gauge Theories: (1+1)-Dimensional Exercises}",
    reportNumber = "CLNS-318",
    doi = "10.1103/PhysRevD.13.1043",
    journal = "Phys. Rev. D",
    volume = "13",
    pages = "1043",
    year = "1976"
}

@article{Susskind:1976jm,
    author = "Susskind, Leonard",
    title = "{Lattice Fermions}",
    reportNumber = "PTENS-76-1",
    doi = "10.1103/PhysRevD.16.3031",
    journal = "Phys. Rev. D",
    volume = "16",
    pages = "3031--3039",
    year = "1977"
}

@article{Li:2024dpq,
    author = "Li, Yang-Yang and Wang, Juven and You, Yi-Zhuang",
    title = "{Quantum Many-Body Lattice C-R-T Symmetry: Fractionalization, Anomaly, and Symmetric Mass Generation}",
    eprint = "2412.19691",
    archivePrefix = "arXiv",
    primaryClass = "cond-mat.str-el",
    month = "12",
    year = "2024"
}

@article{Ginsparg:1981bj,
    author = "Ginsparg, Paul H. and Wilson, Kenneth G.",
    title = "{A Remnant of Chiral Symmetry on the Lattice}",
    reportNumber = "CLNS-81-520, HUTP-81-A060",
    doi = "10.1103/PhysRevD.25.2649",
    journal = "Phys. Rev. D",
    volume = "25",
    pages = "2649",
    year = "1982"
}

@article{Neuberger:1998wv,
    author = "Neuberger, Herbert",
    title = "{More about exactly massless quarks on the lattice}",
    eprint = "hep-lat/9801031",
    archivePrefix = "arXiv",
    reportNumber = "RU-98-03",
    doi = "10.1016/S0370-2693(98)00355-4",
    journal = "Phys. Lett. B",
    volume = "427",
    pages = "353--355",
    year = "1998"
}

@article{Wang:2013yta,
    author = "Wang, Juven and Wen, Xiao-Gang",
    title = "{Nonperturbative regularization of (1+1)-dimensional anomaly-free chiral fermions and bosons: On the equivalence of anomaly matching conditions and boundary gapping rules}",
    eprint = "1307.7480",
    archivePrefix = "arXiv",
    primaryClass = "hep-lat",
    doi = "10.1103/PhysRevB.107.014311",
    journal = "Phys. Rev. B",
    volume = "107",
    number = "1",
    pages = "014311",
    year = "2023"
}

@article{Wang:2018ugf,
    author = "Wang, Juven and Wen, Xiao-Gang",
    title = "{A Solution to the 1+1D Gauged Chiral Fermion Problem}",
    eprint = "1807.05998",
    archivePrefix = "arXiv",
    primaryClass = "hep-lat",
    doi = "10.1103/PhysRevD.99.111501",
    journal = "Phys. Rev. D",
    volume = "99",
    number = "11",
    pages = "111501",
    month = "7",
    year = "2018"
}

@article{Razamat:2020kyf,
    author = "Razamat, Shlomo S. and Tong, David",
    title = "{Gapped Chiral Fermions}",
    eprint = "2009.05037",
    archivePrefix = "arXiv",
    primaryClass = "hep-th",
    doi = "10.1103/PhysRevX.11.011063",
    journal = "Phys. Rev. X",
    volume = "11",
    number = "1",
    pages = "011063",
    year = "2021"
}

@article{Tong:2021phe,
    author = "Tong, David",
    title = "{Comments on symmetric mass generation in 2d and 4d}",
    eprint = "2104.03997",
    archivePrefix = "arXiv",
    primaryClass = "hep-th",
    doi = "10.1007/JHEP07(2022)001",
    journal = "JHEP",
    volume = "07",
    pages = "001",
    year = "2022"
}

@article{Zeng:2022grc,
    author = "Zeng, Meng and Zhu, Zheng and Wang, Juven and You, Yi-Zhuang",
    title = "{Symmetric Mass Generation in the 1+1 Dimensional Chiral Fermion 3-4-5-0 Model}",
    eprint = "2202.12355",
    archivePrefix = "arXiv",
    primaryClass = "cond-mat.str-el",
    doi = "10.1103/PhysRevLett.128.185301",
    journal = "Phys. Rev. Lett.",
    volume = "128",
    number = "18",
    pages = "185301",
    year = "2022"
}

@article{Wang:2022ucy,
    author = "Wang, Juven and You, Yi-Zhuang",
    title = "{Symmetric Mass Generation}",
    eprint = "2204.14271",
    archivePrefix = "arXiv",
    primaryClass = "cond-mat.str-el",
    doi = "10.3390/sym14071475",
    journal = "Symmetry",
    volume = "14",
    number = "7",
    pages = "1475",
    year = "2022"
}

@article{Wang:2022fzc,
    author = "Wang, Juven",
    title = "{CT or P problem and symmetric gapped fermion solution}",
    eprint = "2207.14813",
    archivePrefix = "arXiv",
    primaryClass = "hep-th",
    doi = "10.1103/PhysRevD.106.125007",
    journal = "Phys. Rev. D",
    volume = "106",
    number = "12",
    pages = "125007",
    year = "2022"
}

@article{Onogi:2025tev,
    author = "Onogi, Tetsuya and Wada, Hiroki and Yamaoka, Tatsuya",
    title = "{Discrete symmetry and 't Hooft anomalies for 3450 model}",
    eprint = "2501.18156",
    archivePrefix = "arXiv",
    primaryClass = "hep-lat",
    reportNumber = "OU-HET-1261",
    doi = "10.22323/1.466.0378",
    journal = "PoS",
    volume = "LATTICE2024",
    pages = "378",
    year = "2025"
}

@article{Zohar:2015hwa,
    author = "Zohar, Erez and Cirac, J. Ignacio and Reznik, Benni",
    title = "{Quantum Simulations of Lattice Gauge Theories using Ultracold Atoms in Optical Lattices}",
    eprint = "1503.02312",
    archivePrefix = "arXiv",
    primaryClass = "quant-ph",
    doi = "10.1088/0034-4885/79/1/014401",
    journal = "Rept. Prog. Phys.",
    volume = "79",
    number = "1",
    pages = "014401",
    year = "2016"
}

\end{document}